\documentclass{appolb}
\usepackage{epsfig}
\usepackage{graphicx}

\begin{document}
\title{Exclusive open strangeness production \\in the $pp \to pp K^{+} K^{-}$ reaction at high energies\\
and a measurement of scalar $\chi_{c0}$ meson%
\thanks{Presented at \textit{Strangeness in Quark Matter 2011}, Sept. 18-24, Cracow, Poland.}%
}
\author{Piotr Lebiedowicz
\address{
The H. Niewodnicza\'{n}ski Institute of Nuclear Physics\\
Polish Academy of Sciences\\
ul. Radzikowskiego 152, 31-342 Krak\'ow, Poland}
}
\maketitle
\begin{abstract}
We discuss the results of exclusive
$p p \to p p K^{+} K^{-}$ reaction at high energies which constitutes 
an irreducible background to three-body processes $p p \to p p M$,
where $M=\phi$, $f_{2}(1275)$, $f_{0}(1500)$, $f_{2}'(1525)$, $\chi_{c0}$.
We consider central diffractive contribution
mediated by Pomeron and Reggeon exchanges
including absorption effects
due to proton-proton interaction and kaon-kaon rescattering.
We make predictions for future experiments at RHIC, Tevatron and LHC.
Differential distributions in invariant two-kaon mass, 
kaon rapidities and transverse momenta of kaons are presented.
We discuss a measurement of exclusive production of scalar
$\chi_{c0}$ meson in the proton-(anti)proton collisions via $\chi_{c0} \to K^{+}K^{-}$ decay.
The corresponding amplitude for exclusive
central diffractive $\chi_{c0}$ meson production
is calculated within the $k_{t}$-factorization approach.
\end{abstract}
  
\section{Introduction}

Processes of central exclusive production (CEP)
became recently a very active field of research (\eg Ref. \cite{ACF10}).
Although the attention is paid mainly to high-$p_{t}$ processes
that can be used for new physics searches (exclusive Higgs, $\gamma\gamma$ interactions, etc.),
measurements of low-$p_{t}$ signals are also very important as they can
help to constrain models of the backgrounds for the former ones.
We have studied $pp \to pp \pi^+ \pi^-$ process for low and high energies \cite{SL09,LSK09,LS10}
and $pp \to nn \pi^+ \pi^+$ process at high energies \cite{LS11}.
In Ref. \cite{SLTCS11} a possible measurement of the exclusive $\pi^+ \pi^-$ production
at the LHC with tagged forward protons has been studied.
Recently, we have discussed \cite{LS11_kaons}
the mechanisms of exclusive $K^{+} K^{-}$ production 
in hadron collisions at high energies.
The $p p \to pp K^{+} K^{-}$ reaction is a natural background for exclusive production
of resonances decaying into $K^{+} K^{-}$ channel, such as:
$\phi$, $f_{2}(1270)$, $f_{0}(1500)$, $f_{2}'(1525)$, $\chi_{c0}$.
The mass spectrum of the exclusive $K^{+}K^{-}$ system 
at the CERN Intersecting Storage Rings (ISR)
was measured at $\sqrt{s} = 63$ GeV \cite{AFS85}
and at $\sqrt{s} = 62$ GeV \cite{ABCDHW89}
(this is the highest energy at which normalized experimental data exist). 

Recently there was interest in central exclusive production of 
heavy resonance states (see Refs.~\cite{PST_chic0,PST_chic1,PST_chic2,LKRS10,LKRS11})
where the QCD mechanism is similar to the exclusive production of the Higgs boson.
We recall that the CDF measurement of $\chi_c$ CEP \cite{CDF_chic}
is based on the detection of the decay 
$\chi_c \to J/\psi + \gamma$ with $J/\psi \to \mu^{+}\mu^{-}$ channel. 
At the Tevatron the experimental invariant mass resolution $M(J/\psi + \gamma)$
does not allow a separation of the different $\chi_{cJ}$ states.
Therefore, although the cross section for exclusive $\chi_{c0}$ production obtained within the
$k_{t}$-factorization \cite{PST_chic2} should be the largest among $\chi_{c}$ states,
the higher spin $\chi_{c1}$ and $\chi_{c2}$ states could give similar
contributions to the observed $J/\psi + \gamma$ decay channel, 
because of their much higher branching fractions \cite{PDG}.

The observation of $\chi_{c0}$ CEP via two-body decay channels
is of special interest for studying the dynamics of heavy quarkonia production.
The measurement of exclusive production of $\chi_{c0}$ meson
in proton-(anti)proton collisions via $\chi_{c0} \to \pi^{+} \pi^{-}$ decay
has been already discussed \cite{LPS11}.
Recently we have analyzed a possibility to measure $\chi_{c0}$ via its decay to the $K^+ K^-$ channel.
The branching fraction to this channel is relatively larger for scalar meson
than for the tensor meson \cite{PDG}
($\mathcal{B}( \chi_{c0} \to K^{+} K^{-})  = (0.61 \pm 0.035)\%$,
$\mathcal{B}( \chi_{c2} \to K^{+} K^{-})  = (0.109 \pm 0.008)\%$)
and even absent for the axial meson.
A much smaller cross section for $\chi_{c2}$ production as obtained
from theoretical calculation means that only $\chi_{c0}$ will contribute to the signal.

\section{Background and signal amplitudes}

\begin{figure}[!h]
\centering
a) \includegraphics[width=0.23\textwidth]{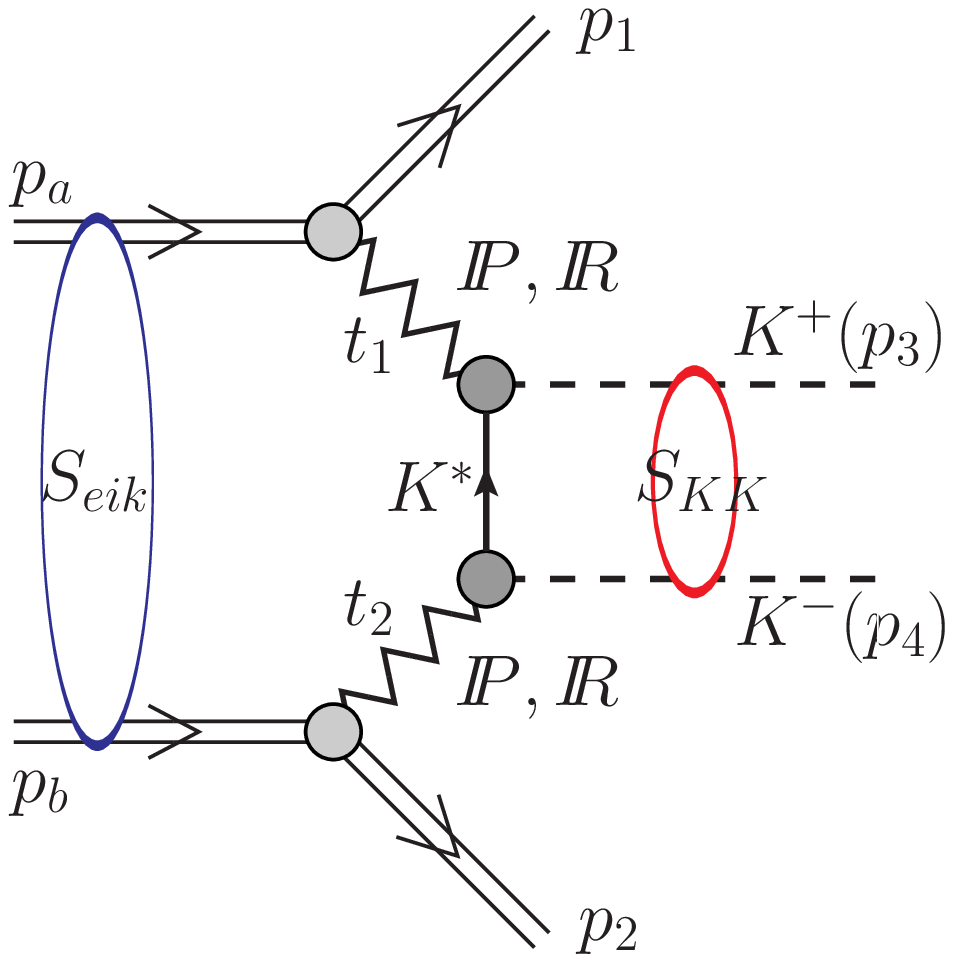}
   \includegraphics[width=0.23\textwidth]{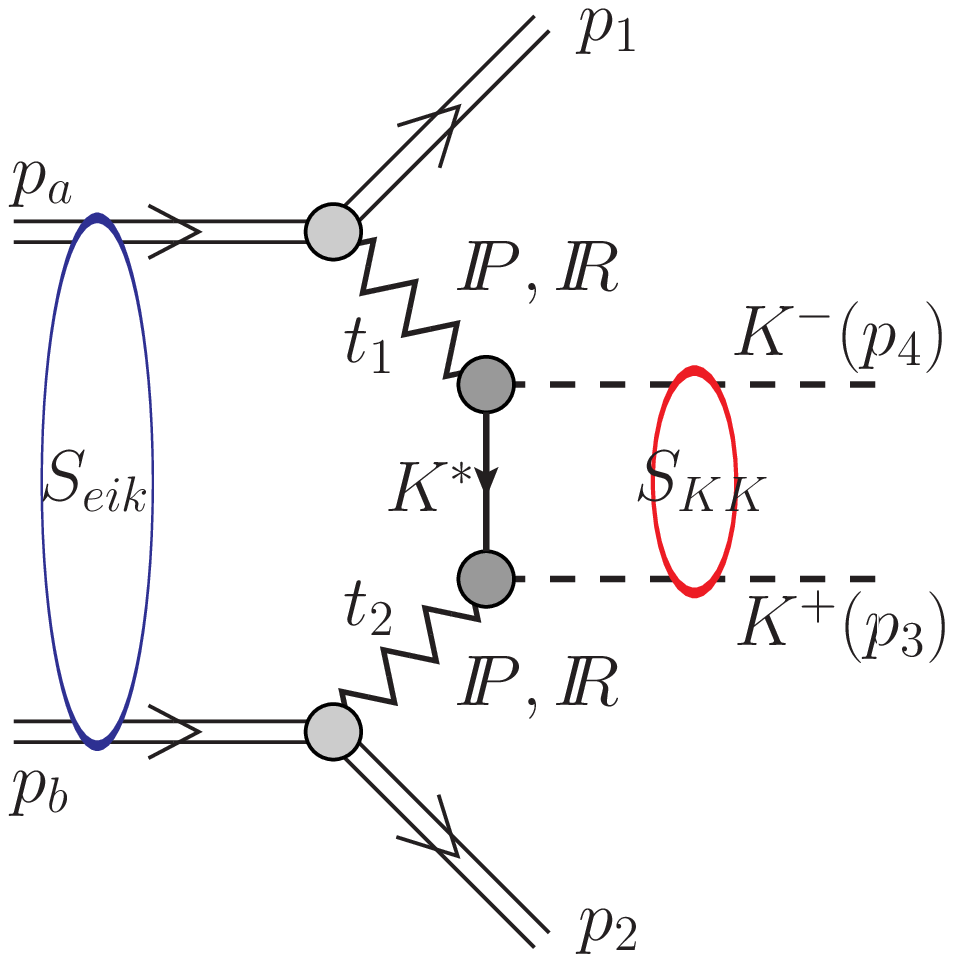} \qquad
b) \includegraphics[width=0.3\textwidth]{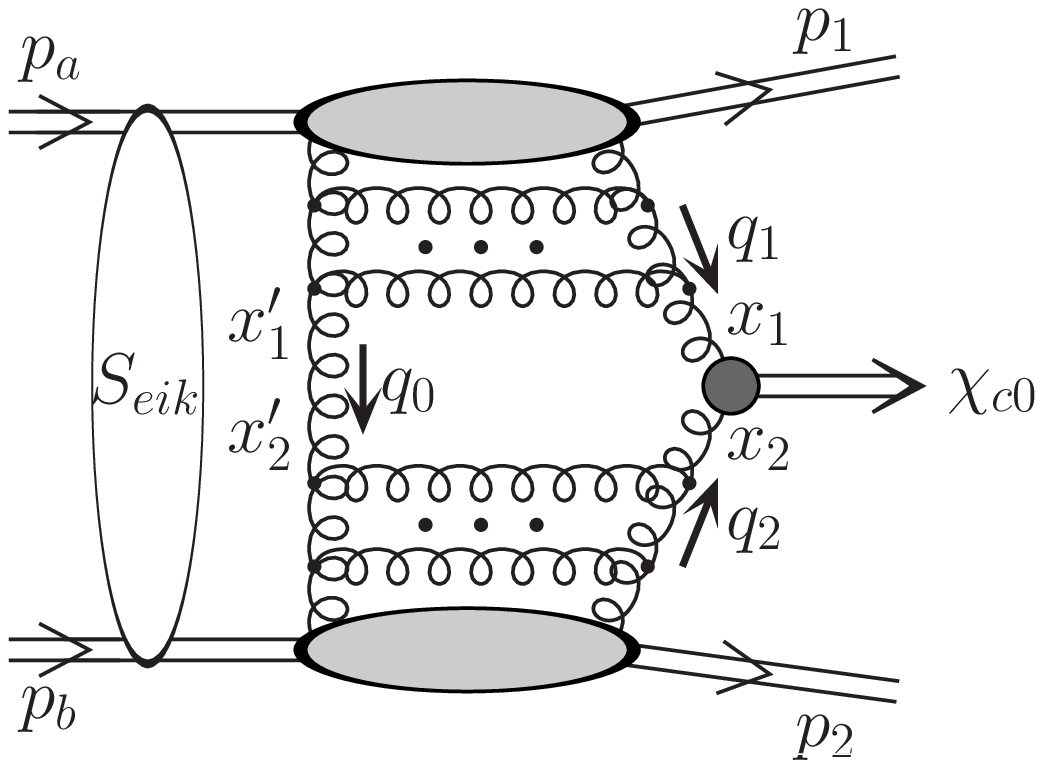}
\caption{
The central diffractive mechanism of exclusive production of $K^{+}K^{-}$ pairs
and the QCD mechanism of $\chi_{c0}$ CEP including absorptive corrections.}
\label{fig:diagrams}
\end{figure}

The dominant mechanism of the exclusive production of
$K^{+}K^{-}$ pairs at high energies is sketched in Fig.\ref{fig:diagrams}(a).
The formalism used in the calculation
of expected non-resonant background amplitude
is explained in Ref.\cite{LS11_kaons}
including the absorptive corrections due to proton-proton interactions
as well as kaon-kaon rescattering.
The Regge parametrization of the $K^{\pm} p \to K^{\pm} p$
scattering amplitude includes both Pomeron and Reggeon exchanges
with the parameters taken from the Donnachie-Landshoff analysis \cite{DL92}
of the total cross sections. 
In Ref.\cite{LS11_kaons} the integrated cross section for the total
and elastic $KN$ scattering was shown.
Our model sufficiently well describes the elastic $KN$ data for energy $\sqrt{s} > 3$ GeV.
The form factors correcting for the off-shellness of the intermediate kaons
are parametrized as
$F_{K}(\hat{t}/\hat{u})=
\exp\left(\frac{\hat{t}/\hat{u}-m_{K}^{2}}{\Lambda^{2}_{off}}\right)$,
where the parameter $\Lambda_{off}^{2} = 2$ GeV$^{2}$ is obtained from a fit 
to the ISR experimental data \cite{ABCDHW89}.

The QCD amplitude for exclusive central diffractive $\chi_{c0}$ meson production,
sketched in Fig.\ref{fig:diagrams}(b),
was calculated within the $k_{t}$-factorization approach
including virtualities of active gluons \cite{PST_chic0}
and the corresponding cross section is calculated with the help of
unintegrated gluon distribution functions (UGDFs).
In Ref.\cite{LPS11} we have performed detailed studies of several differential distributions
of $\chi_{c0}$ meson production.
In the calculation of the $\chi_{c0}$ distributions we have used 
two choices of collinear gluon distributions:
GRV94 NLO \cite{GRV} and GJR08 NLO \cite{GJR}.

\section{Results}

\begin{figure}[!h]
\centering
\includegraphics[width = 0.32\textwidth]{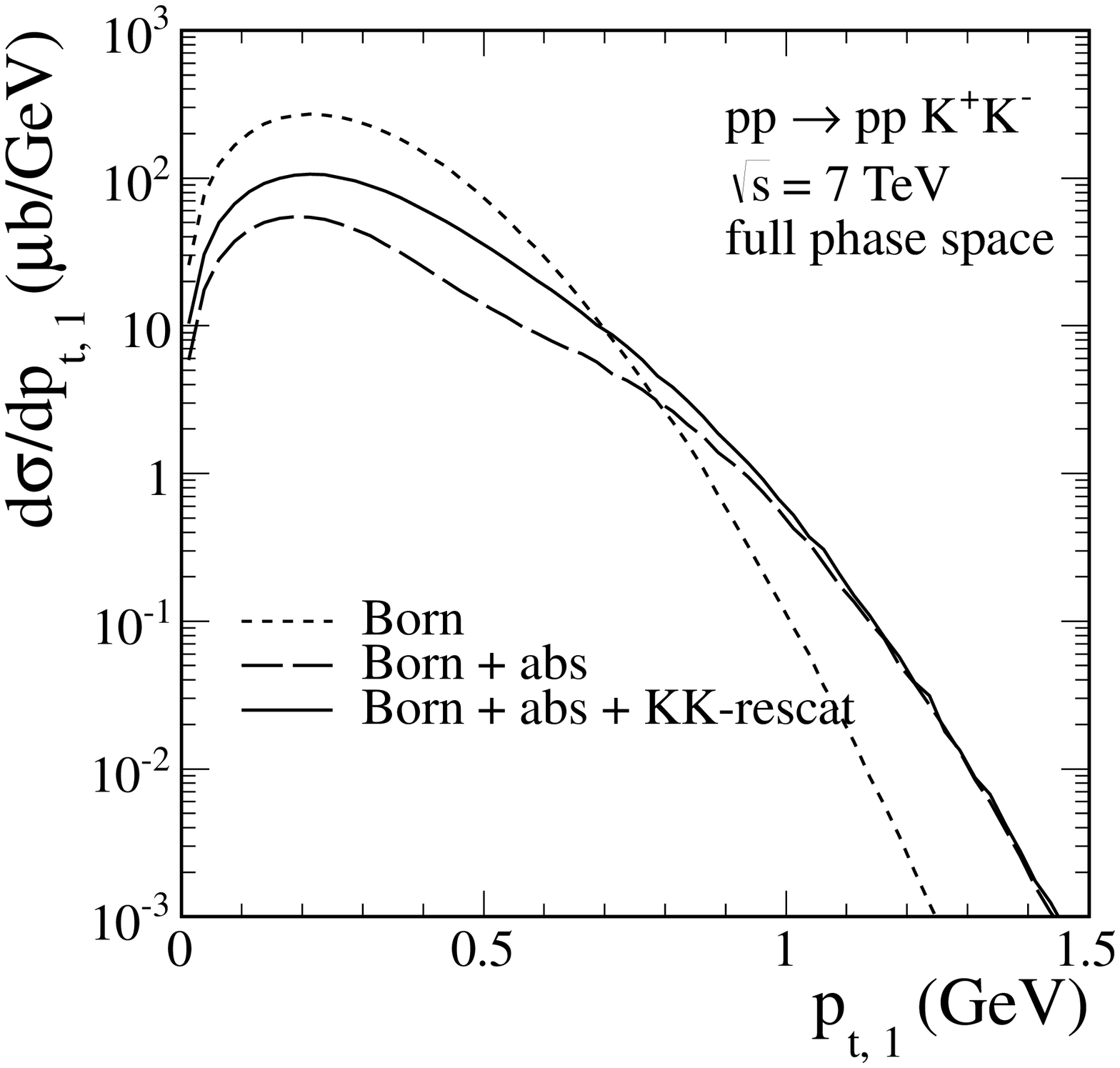}
\includegraphics[width = 0.32\textwidth]{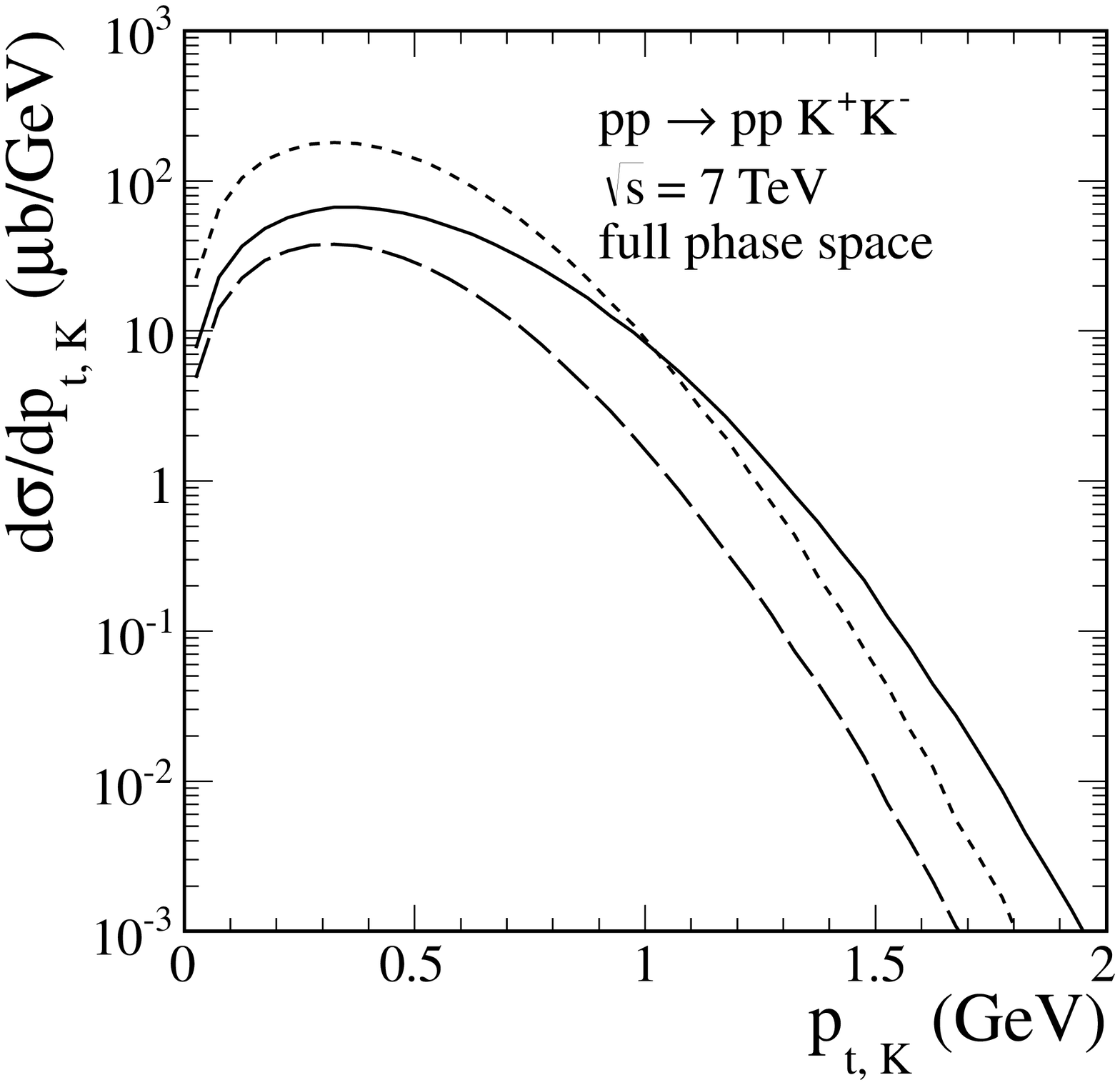}
\includegraphics[width = 0.32\textwidth]{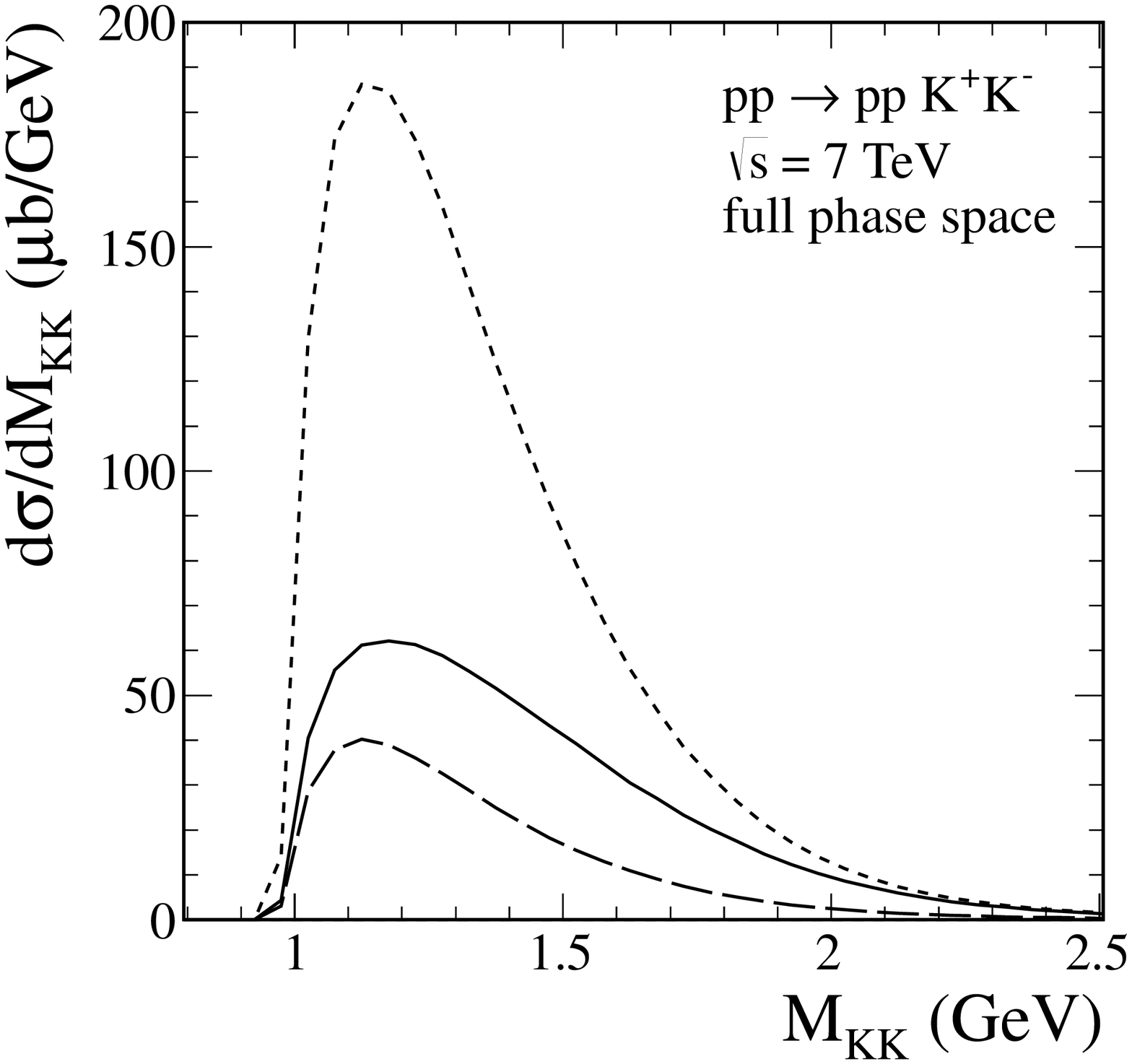}\\
\includegraphics[width = 0.32\textwidth]{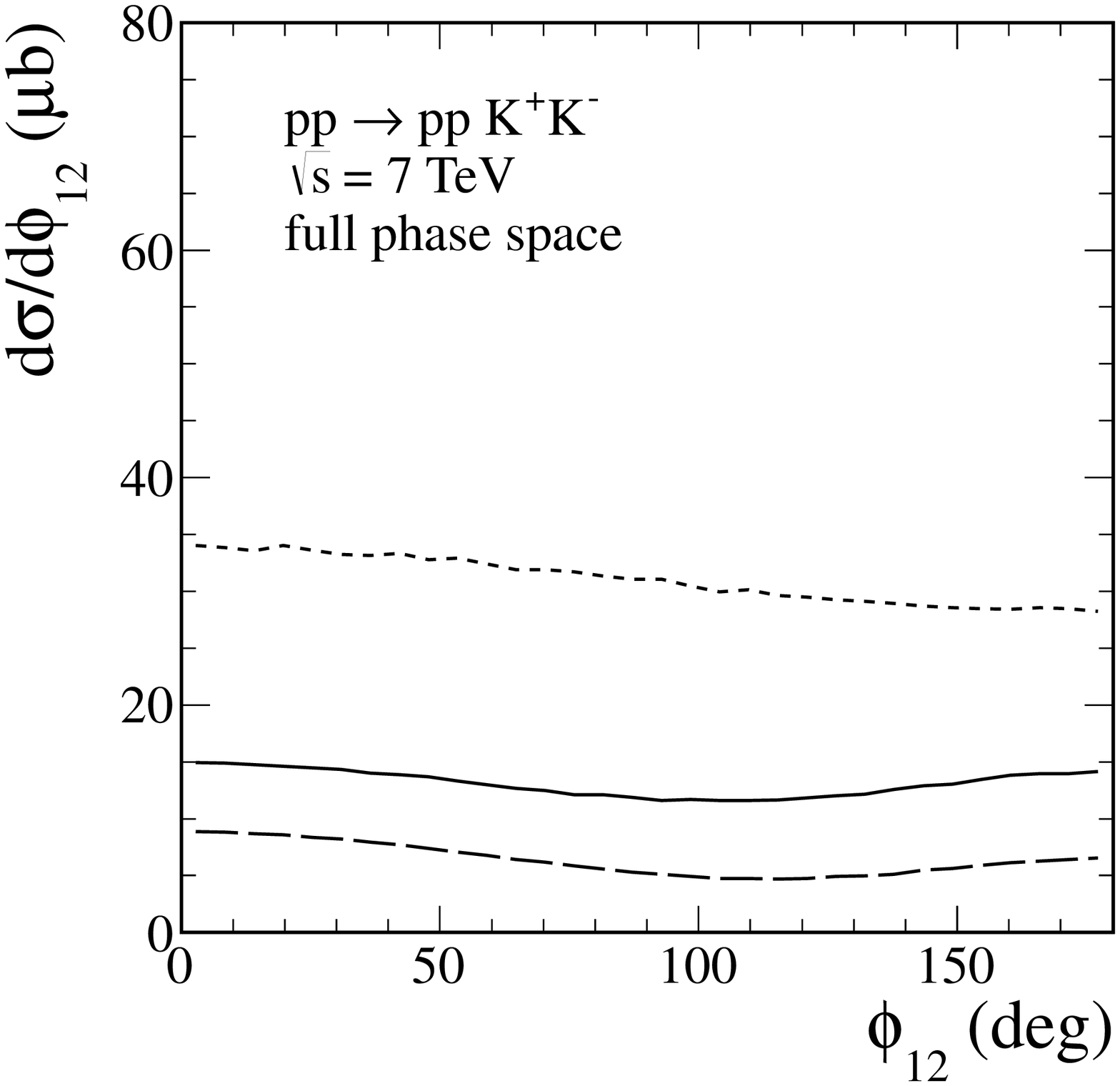}
\includegraphics[width = 0.32\textwidth]{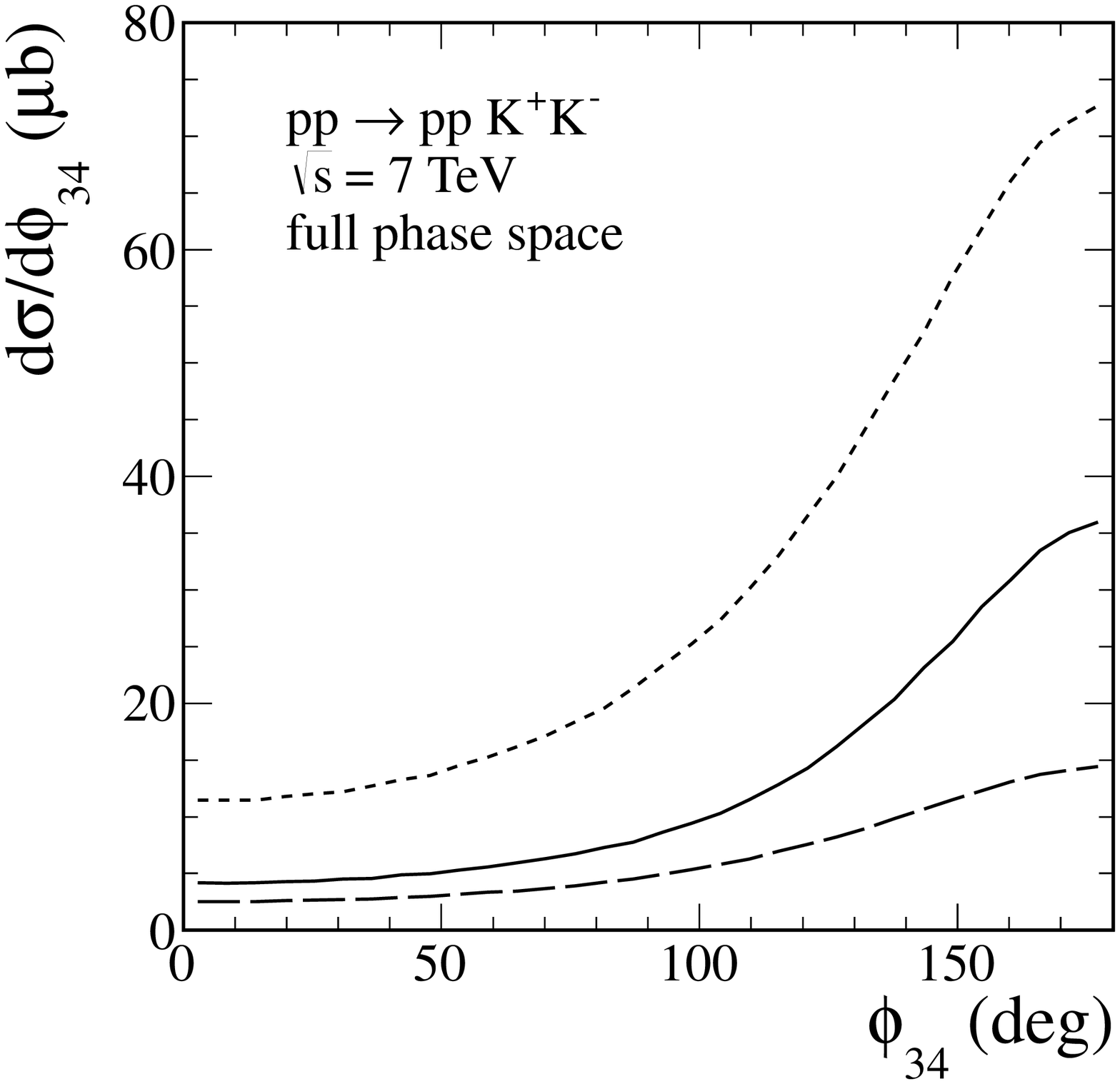}
\includegraphics[width = 0.32\textwidth]{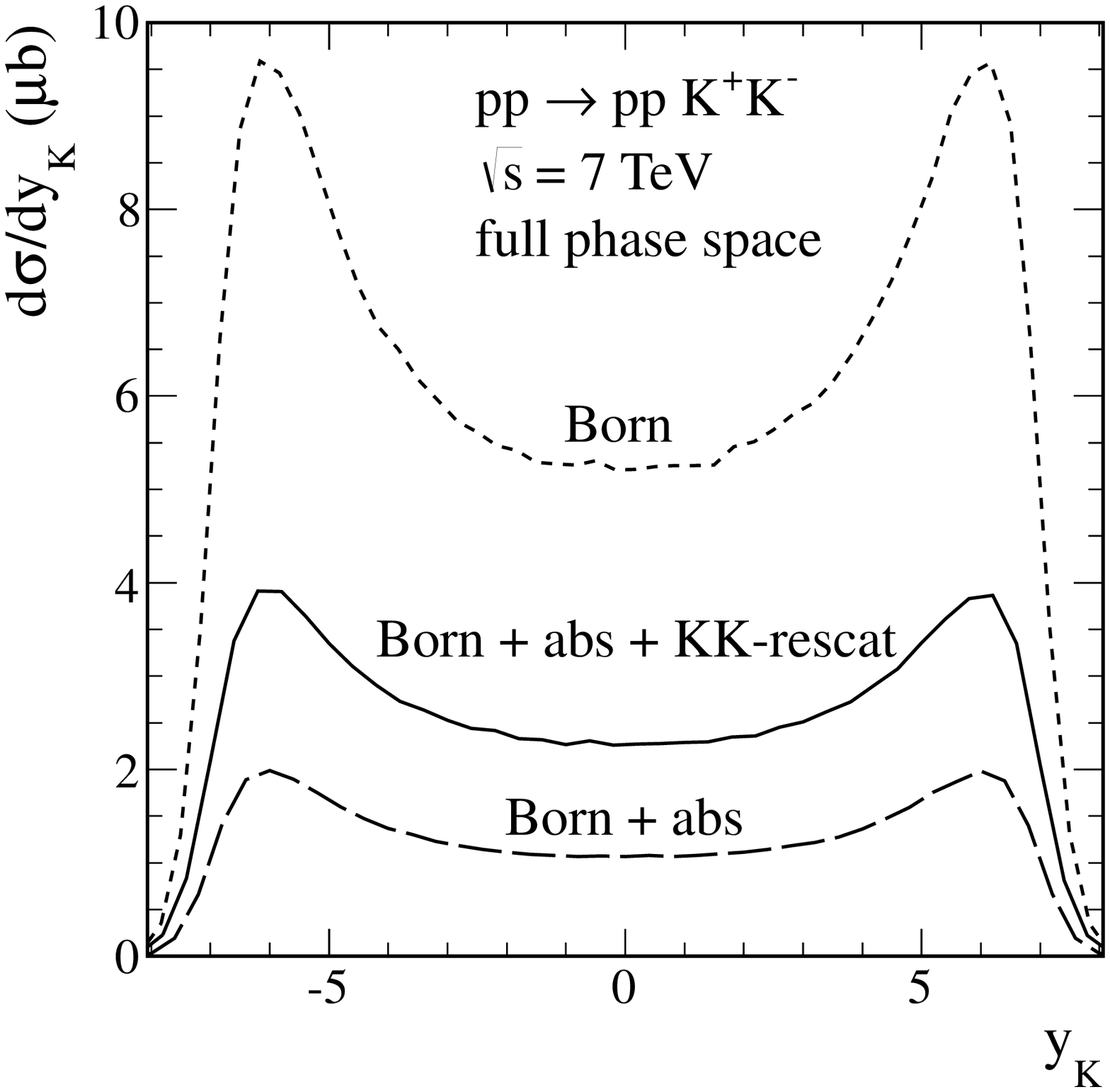}
  \caption{\label{fig:diff_comp}
Differential cross sections for the $pp \to pp K^{+} K^{-}$ reaction
at $\sqrt{s} = 7$ TeV without (dotted line) and with (solid line) the absorption effects.}
\end{figure}

In Fig.~\ref{fig:diff_comp} we show differential distributions
for the $pp \to pp K^{+} K^{-}$ reaction at $\sqrt{s} = 7$ TeV.
In these calculations we include absorption corrections.
In most distributions the shape is almost unchanged.
Our results show that the $KK$-rescattering leads rather
to an enhancement of the cross section 
compared to the calculation without $KK$-rescattering.
The camel-like shape of the rapidity distribution
is due to the interference of different components in the amplitude.
While the Pomeron-Pomeron exchanges peak at midrapidity
the Pomeron-Reggeon (Reggeon-Pomeron) exchanges
peak at backward (forward) kaon rapidities.

\begin{figure}[!h]
\centering
\includegraphics[width = 0.32\textwidth]{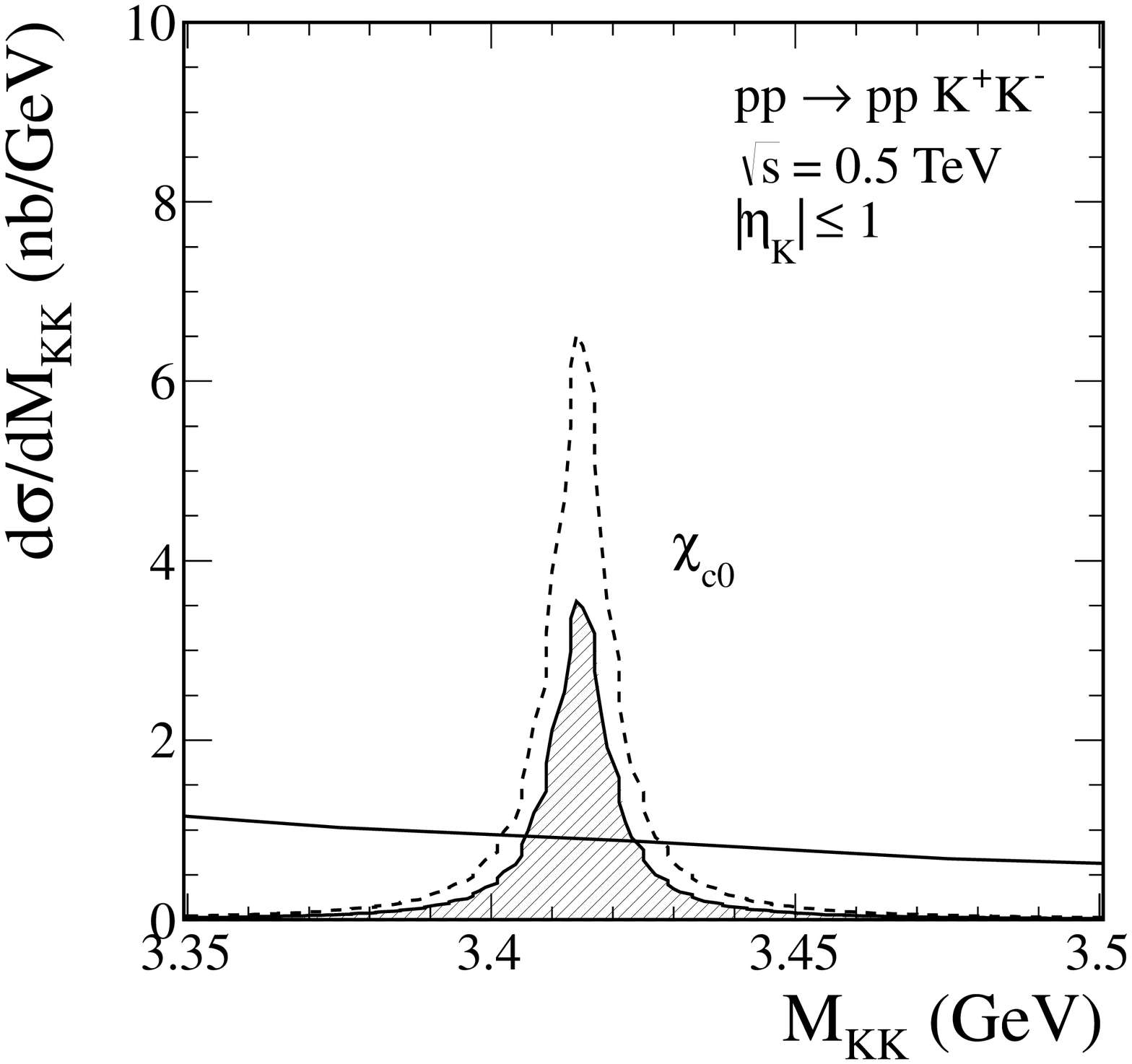}
\includegraphics[width = 0.32\textwidth]{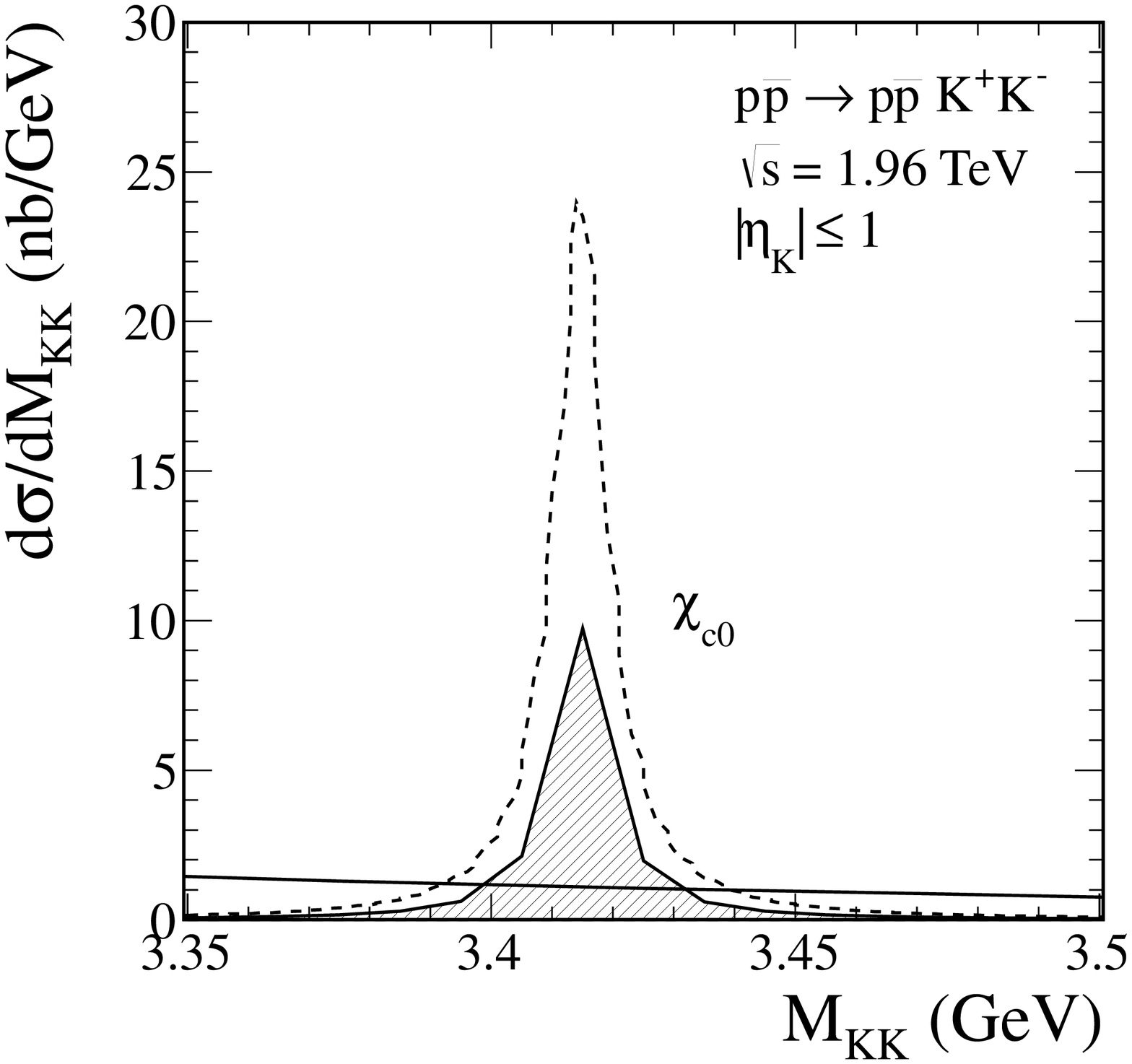}
\includegraphics[width = 0.32\textwidth]{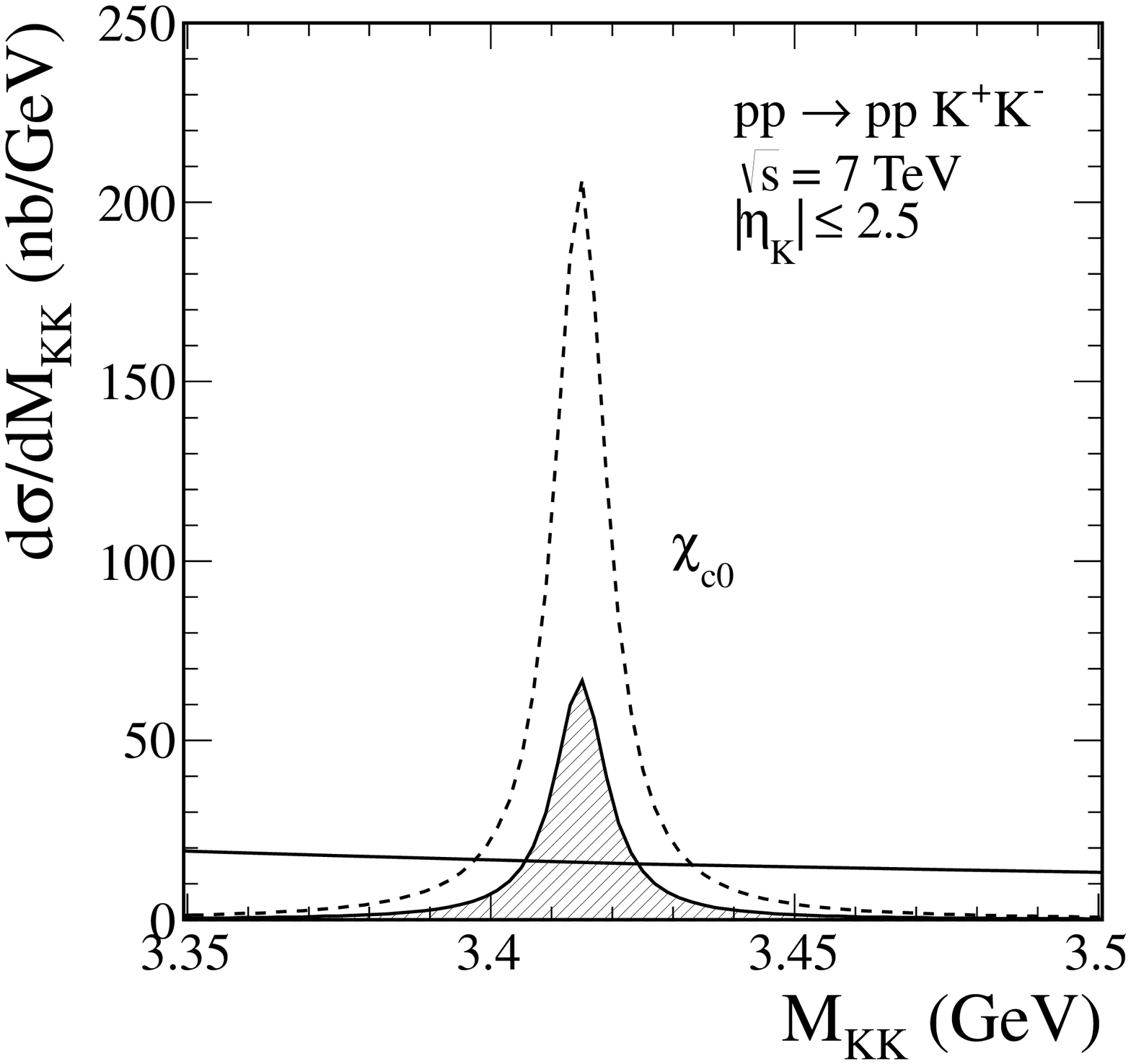}
  \caption{\label{fig:dsig_dmkk}
The $K^{+}K^{-}$ invariant mass distribution at $\sqrt{s} = 0.5,1.96, 7$ TeV 
with the detector limitations in kaon pseudorapidities.
The solid lines present the $KK$-continuum.
The $\chi_{c0}$ contribution
is calculated with the GRV94 NLO (dotted lines) and GJR08 NLO (filled areas) collinear
gluon distributions. 
The absorption effects have been included in the calculations.}
\end{figure}

Now we wish to compare differential distributions of kaon from the
$\chi_{c0}$ decay with those for the continuum.
In Fig.~\ref{fig:dsig_dmkk} we show two-kaon invariant mass
distribution for the central diffractive $KK$ continuum and the
contribution from the decay of the $\chi_{c0}$ meson (see the peak
at $M_{KK} \simeq 3.4$ GeV).
In these figures the resonant $\chi_{c0}$ 
distribution was parameterized in the Breit-Wigner form (see \cite{LS11_kaons}).
Results including the relevant kaon pseudorapidity restrictions 
$-1 < \eta_{K^{+}},\eta_{K^{-}} < 1$ (RHIC and Tevatron) and 
$-2.5 < \eta_{K^{+}},\eta_{K^{-}} < 2.5$ (LHC) are shown.
Clear $\chi_{c0}$ signal with relatively small background can be observed.

\begin{figure}[!h]
\centering
\includegraphics[width = 0.32\textwidth]{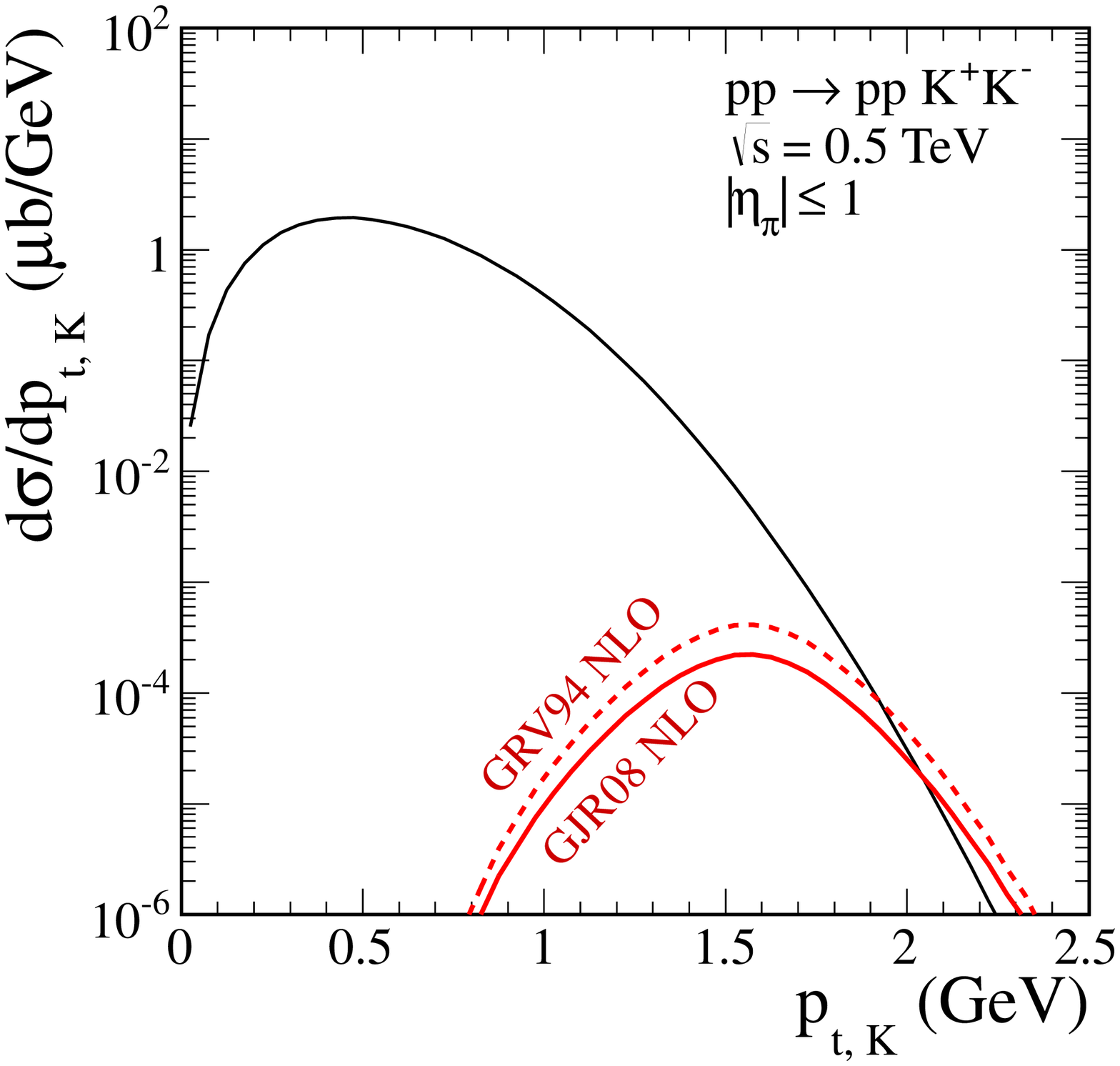}
\includegraphics[width = 0.32\textwidth]{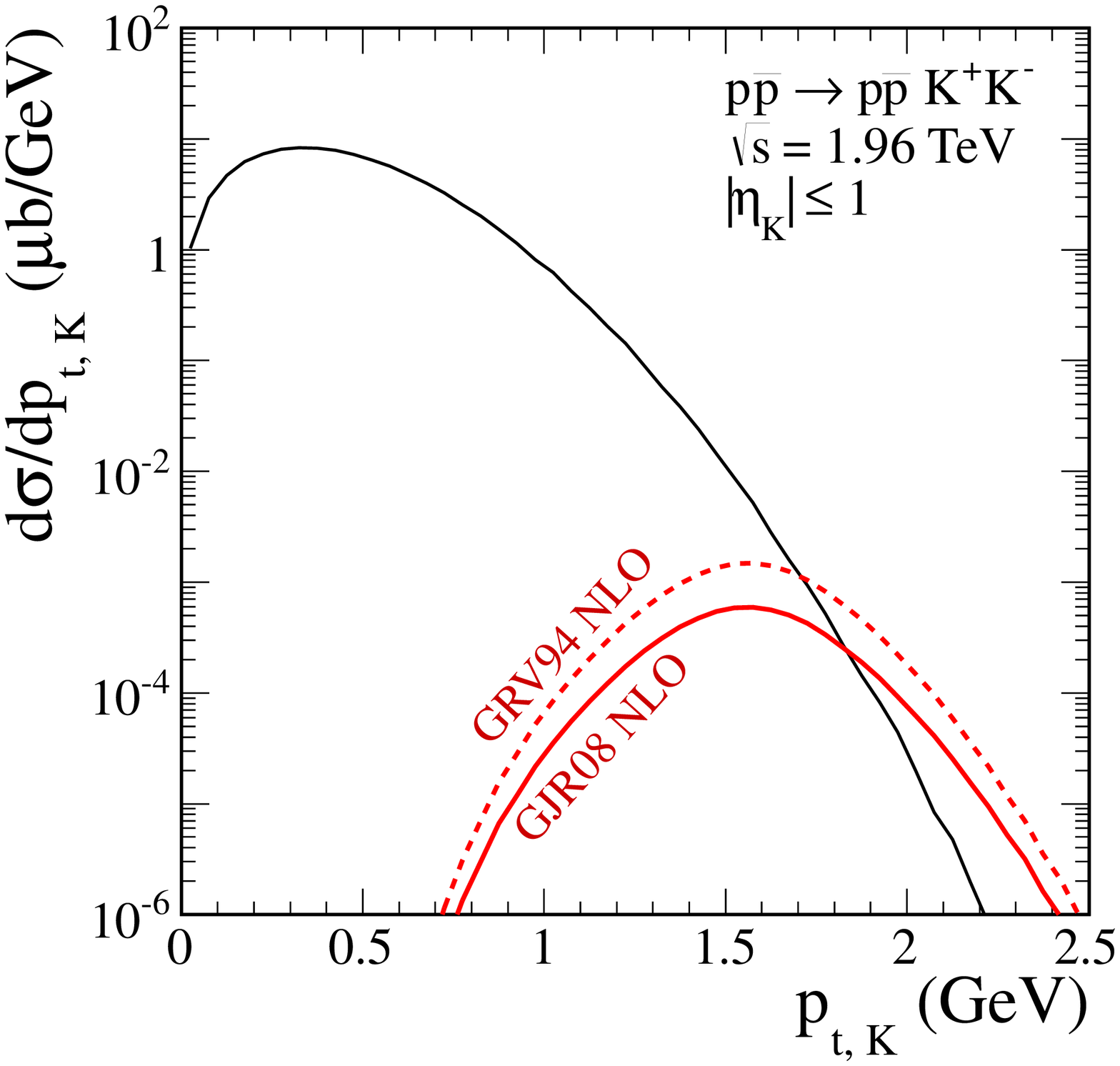}
\includegraphics[width = 0.32\textwidth]{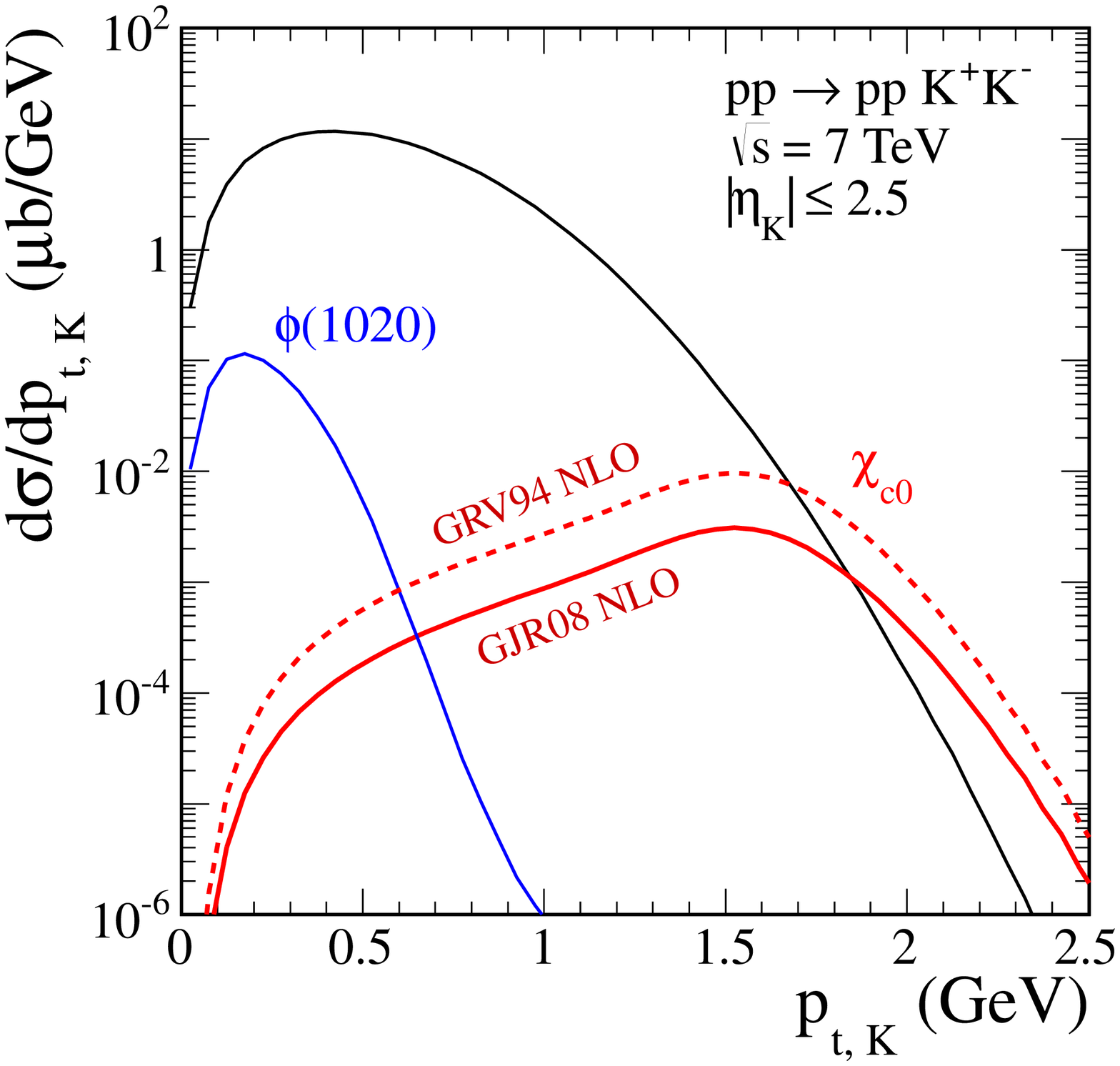}
  \caption{\label{fig:pt_lhc}
Differential cross section $d\sigma/dp_{t,K}$ at $\sqrt{s} = 0.5,
1.96, 7$ TeV with cuts on the kaon pseudorapidities.
Results for the diffractive background (solid lines)
and the kaons from the decay of the $\chi_{c0}$ meson 
including the $K^{+}K^{-}$ branching ratio are shown.
In the right panel $\phi$ meson contribution calculated as in Ref.\cite{CSS10} is shown in addition.
The absorption effects have been included here.}
\end{figure}

In Fig.~\ref{fig:pt_lhc} we show distributions in kaon transverse momenta.
The kaons from the $\chi_{c0}$ decay are placed at slightly larger $p_{t,K}$.
This can be therefore used to get rid of the bulk of the continuum by imposing
an extra cut on the kaon transverse momenta.
It is not the case for the kaons from the $\phi$ meson decay which are placed at lower $p_{t,K}$.

The integrated cross section for exclusive $K^{+}K^{-}$ production 
slowly rises with incident energy, see Table~\ref{tab:sig_tot_kk}.
\begin{table}
\caption{Integrated cross sections 
in $\mu b$ (with absorption corrections) for exclusive $K^{+}K^{-}$ production.
In this calculations we have taken into account
the relevant limitations:
$|\eta_{K}| < 1$ at RHIC and Tevatron,
$|\eta_{K}| < 2.5$ at LHC.
}
\label{tab:sig_tot_kk}
\begin{center}
\begin{tabular}{|c|c|c|}
\hline
$\sqrt{s}$ (TeV) & full phase space & with cuts on $\eta_{K}$\\
\hline
0.5  & 18.47 & 1.21 \\
1.96 & 27.96 & 1.37 \\
7    & 41.14 & 7.38 \\
\hline
\end{tabular}
\end{center}
\end{table}

\section{Conclusions}
We have calculated several differential observables 
for the exclusive $p p \to p p K^+ K^-$ and $p \bar p \to p \bar p K^+ K^-$ reactions.
The full amplitude of central diffractive process
was calculated in a simple model with parameters adjusted to low energy data.
The energy dependence of the amplitudes of the $KN$ subsystems was parametrized in the Regge form
which describes total and elastic cross section for the $KN$ scattering. 
We have predicted large cross sections for RHIC, Tevatron and LHC
which allows to hope that presented by us distributions will be measured in near future.

At the Tevatron the measurement of exclusive production of $\chi_{c}$
via decay in the $J/\psi + \gamma$ channel cannot provide production
cross sections for different species of $\chi_{c}$.
In this decay channel the contributions of $\chi_{c}$ mesons
with different spins are similar and experimental resolution is not
sufficient to distinguish them. However, at LHC situation should be better.

We have analyzed a possibility to measure the
exclusive production of $\chi_{c0}$ meson in the proton-(anti)proton
collisions at the LHC, Tevatron and RHIC via $\chi_{c0} \to K^{+}K^{-}$ decay channel.
We demonstrated how to impose extra cuts in order to
improve the signal-to-background ratio.
For a more detailed discussion of this issue see \cite{LS11_kaons}.
We have shown that relevant measurements at RHIC, Tevatron and LHC are possible
and could provide useful information about the $\chi_{c0}$ exclusive production.

\vspace{0.5cm}
\begin{center}
{\bf Acknowledgments}\\
\end{center}
This work was supported in part by the MNiSW grant
No. PRO-2011/01/N/ST2/04116.

\end{document}